# Inherent correlations between thermodynamics and statistic physics in extensive and nonextensive systems

Zhifu Huang[1], Congjie Ou[2], A. Le Méhauté[2], Qiuping A. Wang[2] and Jincan Chen[1,2,*]

[1]Department of Physics and Institute of Theoretical Physics and Astrophysics, Xiamen University, Xiamen 361005, People's Republic of China

[2]Institut Supérieur des Matériaux et Mécaniques Avancées du Mans, 44 Av. Bartholdi, 72000 Le Mans, France

With the help of a general expression of the entropies in extensive and nonextensive systems, some important relations between thermodynamics and statistical mechanics are revealed through the views of thermodynamics and statistic physics. These relations are proved through the MaxEnt approach once again. It is found that for a reversible isothermal process, the information contained in the first and second laws of thermodynamics and the MaxEnt approach is equivalent. Moreover, these relations are used to derive the probability distribution functions in nonextensive and extensive statistics and calculate the generalized forces of some interesting systems. The results obtained are of universal significance.

*Email: jcchen@xmu.edu.cn

## 1. Introduction

The first and second laws of thermodynamics are two of the most important statements in physics. The probability distribution function of a system is a basic ingredient of the statistical physics. It's a bridge between large numbers of particle movement principles in microscopic view and the thermodynamic laws in macroscopic view, so the hypothesis proposed by Boltzmann and the thermodynamic physics can be unified into a whole. The variation of the probability distribution always affects the thermodynamic properties of a system simultaneously. The probability distribution function determines the number of particles in each energy state [1-3] and the variations of particle number in the microstates lead to the energy exchange between the system and the surroundings. How are they connected by some basic relations? This is a very fundamental problem, which has received increasing attention [4-6]. In this paper, the correlations between the different statistical physics including extensive and nonextensive [7-17] statistics and thermodynamics will be investigated by a unified method and some very general conclusions will be obtained.

## 2. A general expression of the entropy

For different statistical physical methods, the forms of the entropy are very different from each other. We will give a quite general information measure S that, according to Kinchin's axioms for information theory [18-20], depends exclusively on the probability distribution $\{p_i\}$ $(i=1,2,3...W)$. Here a fonctionelle $S(p_i)$ of the probability distribution can be used to unify the different forms of the entropy as

$$S = S(p_i) \qquad (1)$$

where $p_i$ is the probability of the state $i$ among $W$ possible ones that are accessible to the calculation. For different statistical methods, the normalization of the probability distribution is different from each other. However, the normalization of the probability distribution in different statistical mechanics may be unified as

$$\sum_i p_i^{q^\delta} = 1, \qquad (2)$$

where $q \in R$ is a nonextensive parameter, and $\delta = 0$ and 1 for the extensive statistics and Tsallis statistics and for the incomplete statistics, respectively.

## 3. Two important relations between thermodynamics and statistic physics

According to the theory of statistic mechanics, the observable quantities of a physical system can be written as the weighted expectation of the microscopic states, i.e.,

$$\langle O \rangle = \sum_i O_i g(p_i), \qquad (3)$$

where $g(p_i)$ is another fonctionelle of the probability distribution and has different forms for different statistical physical methods. For example, for the Boltzmann-Gibbs statistics,

$$g(p_i) = p_i; \qquad (4a)$$

for the Tsallis statistics,

$$g(p_i) = p_i^q / \sum_j p_j^q; \qquad (4b)$$

for the incomplete statistics,

$$g(p_i) = p_i^q; \qquad (4c)$$

and for other statistics, $g(p_i)$ may have other choices. According to Eq. (3), one has

$$U = \langle \varepsilon_i \rangle = \sum_i g(p_i) \varepsilon_i \qquad (5)$$

and

$$\left\langle \frac{\partial \varepsilon_i}{\partial y_k} \right\rangle = \sum_i g(p_i) \frac{\partial \varepsilon_i}{\partial y_k}, \qquad (6)$$

where $U$ is the internal energy of a system, $\varepsilon_i$ is the energy of the system at state $i$, $y_k$ are the k-th generalized coordinate of the system. Obviously, the energy $\varepsilon_i$ should be a function of the generalized coordinates, i.e.,

$$\varepsilon_i = \varepsilon_i(y_j). \qquad (7)$$

Thus, the probability distribution of the system at temperature $T$ depends on the external parameters, i.e.,

$$p_i = p_i(T, \varepsilon_i(y_j)) \ . \tag{8}$$

From Eqs. (5) and (6), we obtain

$$dU = \sum_{i,j} \frac{\partial g(p_i)}{\partial p_j} \varepsilon_i dp_j + \sum_{i,k} \frac{\partial \varepsilon_i}{\partial y_k} g(p_i) dy_k$$

$$= \sum_{i,j} \frac{\partial g(p_i)}{\partial p_j} \varepsilon_i dp_j + \sum_k \left\langle \frac{\partial \varepsilon_i}{\partial y_k} \right\rangle dy_k \tag{9}$$

which is just the expression of the internal energy derived from the view of statistic mechanics.

On the other hand, according to the first and second laws of thermodynamics,

$$dU = TdS + dW = TdS - \sum_k Y_k dy_k \ , \tag{10}$$

and Eq.(1), we obtain

$$dU = T\sum_j \frac{\partial S}{\partial p_j} dp_j - \sum_k Y_k dy_k, \tag{11}$$

where $Y_k$ are the conjugate generalized forces of the generalized coordinates $y_k$. Comparing Eq. (11) with Eq. (9), we obtain two important relations as follows

$$T\sum_j \frac{\partial S}{\partial P_j} dp_j = \sum_{i,j} \frac{\partial g}{\partial p_j} \varepsilon_i dp_j \tag{12}$$

and

$$Y_k = -\left\langle \frac{\partial \varepsilon_i}{\partial y_k} \right\rangle = -\sum_i \frac{\partial \varepsilon_i}{\partial y_k} g(p_i) \ . \tag{13}$$

It is very significant to note the fact that Eqs. (12) and (13) are true for any method of the statistic mechanics.

On account of the different normalization conditions of the probability distribution, the changes in $p_i$ must satisfy the following relation

$$\sum_j dp_j^{q^\delta} = 0. \tag{14}$$

Using Eqs. (12) and (14), we get

$$T\frac{\partial S(p_i)}{\partial P_j} - \sum_i \frac{\partial g(p_i)}{\partial p_j} \varepsilon_i = Kq^\delta p_j^{q^\delta - 1}, \tag{15}$$

where $K$ is a constant. It should be pointed out that Eq. (15) is an essential generalization of Eq. (31) in Ref. [5] as well as the result obtained in Ref. [4]. It is suitable not only for extensive systems but also for nonextensive systems described by the Tsallis statistics, incomplete statistics, or other statistic mechanics.

## 4. The MaxEnt approach

In order to better understand the meaning of Eqs. (13) and (15), one can maximize the entropy of a system which is subjected to some constraint conditions. When there is not any work exchange between the system and the surroundings, the objective function may be given by [6]

$$L_U = S/k - \alpha \sum_i p_i^{q^\delta} - \beta U \,. \tag{16}$$

Equation (16) has been used to calculate the probability distribution when the entropy attains its maximum value under the given temperature condition [4-10]. It is significant to note that when there does not exist the work exchange between the system and the surroundings, it is equivalent to take the internal energy $U$ of the system as one constraint condition and choose the heat exchange $Q$ between the system and the surroundings as the other constraint condition.

When there exists the work exchange between the system and the surroundings in a reversible isothermal process, it is not enough to have only two constraint conditions appearing in the second and third terms on the right hand side of Eq.(16) in order to extremize S. It needs an additional constraint condition, which should be that the exchanged work $W$ in the reversible isothermal process is fixed. According to the first law of thermodynamics, $U + W = Q$, the heat exchange $Q$ between the system and the surroundings in such a process with the reversible work exchange should be a fixed quantity. Thus, one can write a new objective function as

$$L_Q = S/k - \alpha \sum_i p_i^{q^\delta} - \beta Q = S/k - \alpha \sum_i p_i^{q^\delta} - \beta(U + W) \,. \tag{17}$$

It is significant to note the fact that for two different cases without and with the work exchange described by Eqs. (16) and (17), the constraint conditions to maximize the entropy of a system are essentially the same, i.e., both $\sum_i p_i^{q^\delta}$ and $Q$ under the given temperature condition are fixed.

When temperature is fixed, the variance of Eq. (17) leads to

$$\frac{1}{k}\sum_j \frac{\partial S}{\partial p_j}dp_j - \sum_j \alpha q^\delta p_j^{q^\delta-1}dp_j - \beta[\sum_{i,j}\frac{\partial g(p_i)}{\partial p_j}\varepsilon_i dp_j + \sum_{i,k}\frac{\partial \varepsilon_i}{\partial y_k}g(p_i)dy_k + \sum_k Y_k dy_k] = 0. \quad (18)$$

From Eq.(18), one has

$$\sum_j [\frac{1}{k}\frac{\partial S(p_i)}{\partial p_j} - \alpha q^\delta p_j^{q^\delta-1} - \beta \sum_i \frac{\partial g(p_i)}{\partial p_j}\varepsilon_i]dp_j = 0 \quad (19)$$

and

$$\sum_k [\sum_i \frac{\partial \varepsilon_i}{\partial y_k}g(p_i) + Y_k]dy_k = 0. \quad (20)$$

Letting $\alpha/\beta = K$, we can directly derive Eqs. (15) and (13) from Eqs.(19) and (20) once again. It implies the fact that for a reversible isothermal process, the information contained in the first and second laws of thermodynamics, $dU = TdS - \sum_i Y_i dy_i$, and the MaxEnt approach is equivalent.

## 5. The probability distribution of a system

It is seen from Eq.(15) that as long as the forms of $S(p_i)$ and $g(p_i)$ are known, one can conveniently derive the probability distribution of a system from Eq.(15).

Example 1. Tsallis' statistics [7, 8].

Using Tsallis' entropy

$$S = k\frac{\sum_i p_i^q - 1}{1-q} \quad (21)$$

and Eqs. (4b) and (15), one can derive the probability distribution of a system satisfying Tsallis' statistics as [6, 8]

$$p_i = \frac{1}{Z}\left[1 - (1-q)\beta(\varepsilon_i - U)/\sum_j p_j^q\right]^{1/(1-q)}, \quad (22)$$

where

$$Z = \sum_i \left[1-(1-q)\beta(\varepsilon_i - U)/\sum_j p_j^q\right]^{1/(1-q)}.$$

Example 2. Incomplete statistics [9, 10].

Using the expression of the entropy in the incomplete statistics

$$S = k\frac{\sum_i p_i - 1}{q-1} \qquad (23)$$

and Eqs. (4c) and (15), one can derive the probability distribution of a system satisfying the incomplete statistics as [21]

$$p_i = \frac{1}{Z}\left[1-(1-q)q\beta(\varepsilon_i - U)/\sum_j p_j\right]^{1/(1-q)}, \qquad (24)$$

where

$$Z = \left\{\sum_{i=1}^w [1-(1-q)q\beta(\varepsilon_i - U_q)/\sum_{j=1}^w p_j]^{\frac{q}{1-q}}\right\}^{1/q}.$$

Obviously, when $q=1$, both the Tsallis statistics and the incomplete statistics become the Boltzmann-Gibbs statistics. Thus, the probability distribution of a system satisfying the Boltzmann-Gibbs statistics

$$p_i = \frac{1}{Z}\exp(-\beta\varepsilon_i) \qquad (25)$$

can be directly derived from Eq. (22) or (24), where $Z = \sum_i \exp(-\beta\varepsilon_i)$.

## 6. The generalized forces of a system

It is seen from Eq. (13) that as long as the forms of $\varepsilon_i(y_k)$ and $g(p_i)$ are known, one can conveniently derive the generalized forces of a system from Eq. (13). We will discuss several interesting cases below.

If there is $\frac{\partial \varepsilon_i}{\partial y_k} = 0$ for a system, we can obtain the generalized force of the system

$$Y_k = 0 \qquad (26)$$

from Eq.(13). The free particles just belong to such a case.

If $\dfrac{\partial \varepsilon_i}{\partial y_k} = -c(y_i)$, we can obtain

$$Y_k = c(y_k) \tag{27}$$

from Eq.(13). The particles in the gravity field just belong to such a case, i.e., $c(y_k) = -mg$, where $g$ is the gravity constant.

If $\dfrac{\partial \varepsilon_i}{\partial y_k} = -c(y_i)\varepsilon_i$, we can obtain

$$Y_k = c(y_k)U \tag{28}$$

from Eq.(13). The ideal quantum gases [22] confined an n-dimension infinite trap just belong to such a case. For example, for one dimension case,

$$\varepsilon_i = \frac{h}{8m}\frac{n_i^2}{L^2}, \qquad (n_i = 1,2,3,...) \tag{29}$$

and

$$c(L) = 2/L, \tag{30}$$

where $h$ is the Planck constant, $m$ is the mass of a particle, and $L$ is the width of the trap. In such a case, the generalized force corresponds to the pressure $P$ of the system. Substituting Eq. (30) into Eq. (28), we obtain an important relation as

$$PL = 2U. \tag{31}$$

For three dimension case,

$$\varepsilon_i = \frac{h}{8m}\left(\frac{n_i^2}{L_x^2} + \frac{n_j^2}{L_y^2} + \frac{n_k^2}{L_z^2}\right) = \varepsilon_x + \varepsilon_y + \varepsilon_z, \quad (n_i, n_j, n_k = 1,2,3,...), \tag{32}$$

$$c(L_k) = 2/L_k, \quad (k = x, y, z), \tag{33}$$

and

$$Y_k = c(L_k)U_k = \frac{2}{L_k}U_k, \tag{34}$$

where $U_k = \sum_i \varepsilon_x(n_i)g(p_i)$. In such a case, the productions $Y_k y_k$ of the generalized forces and the generalized coordinates correspond to $P_k V$, where $P_k$ is the component of the pressure tensor along direction $k$ and $V$ is the volume of the system. Using Eq. (34), we obtain an important relation between the three components of the pressure tensor and the internal energy of the ideal quantum gases confined in a rectangular box with the side lengths $L_k$ as

$$(P_x + P_y + P_z)V = 2U. \qquad (35)$$

It can be seen Eqs. (34) and (35) that for the ideal quantum gases in a confined space, the pressure tensor of the system is, in general, no longer isotropic because of the geometry effect of the boundary. For n-dimension case, one may further prove that there exists an important relation between the n components of the pressure tensor and the internal energy of the confined quantum gases as

$$\sum_{k=1}^{n} P_k V_n = 2U, \qquad (36)$$

where $V_n$ is the n dimension volume of the system.

If the form of $\varepsilon_i(y_k)$ is more complex than the cases mentioned above, one may calculate the generalized forces of a system according to the concrete forms of $\varepsilon_i(y_k)$ and $g(p_i)$.

## 7. Conclusions

We have derived the important relations between thermodynamics and statistical mechanics in extensive and nonextensive systems through two different approaches and revealed the inherent correlations between thermodynamics and statistical physics. We have proved the equivalence between the information contained in the first and second laws of thermodynamics and the MaxEnt approach for a reversible isothermal process. It is important to generalize essentially the conclusions obtained in Refs. [4, 5] and confirm to Abe's standpoint [13], i.e., statistical mechanics may be modified but thermodynamics should remain unchanged. It is more important that the results obtained here are universal. They are suitable not only for extensive systems but also for nonextensive systems.

**ACKNOWLEDGLENTS**

This work was supported by the National Natural Science Foundation (No. 10875100), People's Republic of China.